\documentstyle[amssymb,prl,aps,epsfig,multicol]{revtex}

\begin{document}

\title{Evidence for three dimensional superconductivity in the
cuprates oxides}
\author{T. Schneider}
\address{Physik-Institut der Universit\"{a}t Z\"{u}rich,
Winterthurerstrasse 190, CH-8057 Z\"{u}rich, Switzerland}

\maketitle

\begin{abstract}
We review the empirical scenario emerging from the measured doping
dependence of the transition temperature and the anisotropy parameter $%
\gamma $ $=\xi _{ab}/\xi _{c}$, defined as the ratio of the
correlation lengths parallel and perpendicular to the ab-planes.
It suggests that two dimensional models cannot explain the
occurrence of superconductivity in the cuprates. This conclusion
is confirmed and extended in terms of a novel scaling relation. It
involves the transition temperature, the aerial superfluid density
in the ground state, $\gamma $ and the dynamic critical exponent
of the quantum superconductor to insulator transition. The
important implication there is that a non vanishing superfluid
density in the ground state of the cuprates is unalterably linked
to an anisotropic but 3D condensation mechanism.
\end{abstract}

\bigskip
\bigskip
\bigskip

\begin{multicols}{2}
\narrowtext

\begin{figure} \centering
\includegraphics[width= 0.95\linewidth]{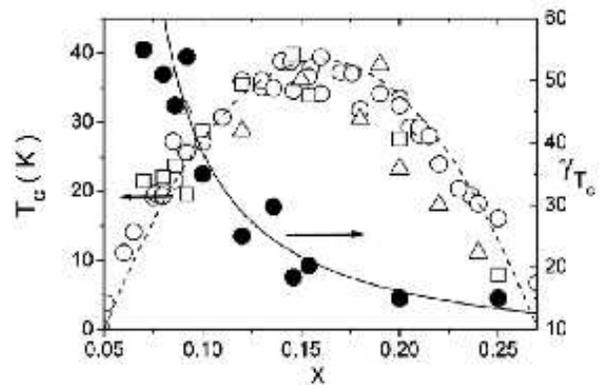} \vskip12pt
\caption{$T_{c}$ ($\bigcirc $)and $\protect\gamma _{T_{c}}$
($\bullet $) versus $x$ for La$_{2-x}$
Sr$_{x}$CuO$_{4}$\protect\cite
{takagi,torrance,suzuki,kimura,yamada,nagano,fukuzumi,sasagawa,hoferdis}.
The dashed curve corresponds to the empirical relation (\ref{eq1}) with $%
T_{c}^{m}=39\ K$, while the solid one is given by Eqs.(\ref{eq1})
and (\ref {eq2}) with $a_{T_{c}\protect\gamma }=5909\ K$. The open
squares follow from Eq.(\ref{eq2}) and the measured
$\protect\gamma _{T_{c}}$, while the open triangles result from
Eq.(\ref{eq3}) and the experimental data for $\protect\gamma
_{T=0}$.} \label{fig1}
\end{figure}

It has been a long-standing question wether two dimensional (2D)
models alone can describe and explain the essential experimental
observations of superconductivity in cuprates oxides. On the one
hand models of the $CuO_{2}$ plane alone, such as the t-J or
Hubbard model, have been successful in explaining properties of
the undoped insulator, including the renormalization of charge
dynamics\cite{anderson,shen} and the occurrence of a spin gap
\cite{fukuyama,kotilar}. Moreover, given the crystal structures of
the superconducting cuprates oxides, it is to be expected that the
physical properties of single crystals will show a high degree of
anisotropy. This is indeed the case; by way of example the ratio
$\rho _{c}/\rho _{ab}$ between the out-of-plane $\rho _{c}$ and
in-plane $\rho _{ab}$ resistivity is very large\cite{semba}. On
the other hand, serious questions can be raised. The first is the
observation that when the ground state is a superconductor both
components of resistivity drop to zero at the same temperature
$T_{c}$, so that the phase transition has genuine 3D
character\cite{wang,semba}. The second emerges from the empirical
phase diagram of La$_{2-x}$Sr$_{x}$CuO$_{4}$\cite
{takagi,torrance,suzuki,kimura,yamada,nagano,fukuzumi,sasagawa,hoferdis}
and
the doping dependence of the anisotropy parameter $\gamma $, depicted in Fig.%
\ref{fig1}. It shows that after passing the so called underdoped limit, at $%
x_{u}\approx 0.05$, where the quantum insulator to superconductor
(QSI) transition occurs \cite{book,klosters,housten,tshk,tsprb},
$T_{c}$ rises and reaches a maximum value $T_{c}^{m}$ at
$x_{m}\approx 0.16$. With further increase of $x$, $T_{c}$
decreases and finally vanishes in the overdoped limit
$x_{o}\approx 0.27$. Here the system undergoes a quantum
superconductor to normal state (QSN) transition. From
Fig.\ref{fig1} it is also seen that decreasing dopant
concentration is also accompanied by a
raise of anisotropy. In tetragonal cuprates it is defined as the ratio $%
\gamma =\xi _{ab}/\xi _{c}$, of the correlation lengths parallel
and perpendicular to the ab-planes. Although the fundamental
reason for the anisotropy increase is not clear, it reveals a
3D-2D crossover with reduced dopant concentration. Note that the
limit $\gamma \rightarrow \infty $
implies 2D critical behavior. The observation that in the underdoped regime $%
T_{c}$ falls with increasing anisotropy $\gamma $ (see
Fig.\ref{fig1}) suggests that superconductivity in the cuprates is
inevitable a 3D phenomenon.

This letter addresses these issues by providing experimental and
theoretical evidence that a non vanishing superfluid density in
the ground state of the cuprates oxides is unalterably linked to
an anisotropic but 3D condensation mechanism. We review the
scenario emerging from the experimental data for the doping
dependence of $T_{c}$ and anisotropy $\gamma $. It suggests that
two dimensional models cannot explain the occurrence of
superconductivity in the cuprates. This conclusion is confirmed
and extended in terms of a new scaling relation. It involves
$T_{c}$, the aerial superfluid density in the ground state,
$\gamma $ and the dynamic critical exponent of the 2D-QSI
transition. The important implication there is that a non
vanishing superfluid density in the ground state of the cuprates
is unalterably linked to an anisotropic but 3D condensation
mechanism. This finding clearly reveals that two dimensional
models cannot explain the occurrence of superconductivity in these
materials.

A generic property of cuprate superconductors is the existence of
a phase transition line, $T_{c}\left( x\right) $, separating the
superconducting from the normal conducting phase, as well as the
so-called underdoped and overdoped regimes. This behavior is
thought to be universal for cuprate superconductors\cite{tallon}.
A glance to Fig.\ref{fig1} reveals that it is very well described
by the empirical relation
\begin{equation}
T_{c}\left( x\right) =T_{c}^{m}(1+82.6(x-0.16)^{2}),  \label{eq1}
\end{equation}
due to Presland \emph{et al.} \cite{presland}. In practice, there
are only a few compounds, including La$_{2-x}$Sr$_{x}$CuO$_{4}$,
HgBa$_{2}$CuO$_{4+x}$ \cite{fukuokahg,hoferhg} and
Bi$_{2}$Sr$_{2}$CuO$_{6+x}$ \cite{groen}, for which the dopant
concentration $x$ can be varied continuously throughout the
entire doping range. In other cuprates, including Bi$_{2}$Sr$_{2}$CaCu$_{2}$O%
$_{8+x}$ \cite{groenbi}, YBa$_{2}$Cu$_{3}$O$_{7-x}$ \cite{semba} and Y$%
_{1-x}\Pr_{x}$Ba$_{2}$Cu$_{3}$O$_{7-\delta }$ \cite{neumeierpr},
only the underdoped and optimally doped regimes appear to be
accessible.
\begin{figure}
\centering
\includegraphics[width=0.95\linewidth]{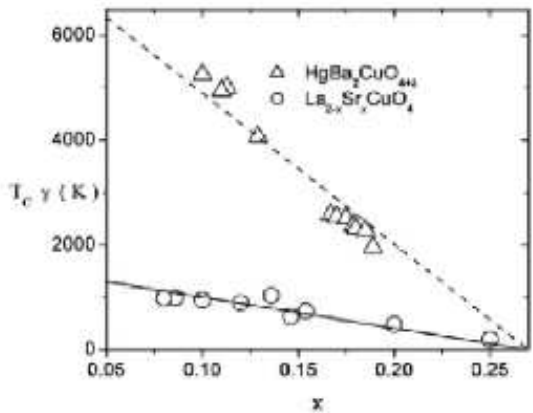} \vskip12pt
\caption{$\ T_{c}\protect\gamma _{T_{c}}$ versus dopant
concentration $x$ for La$_{2-x}$Sr$_{x}$CuO$_{4}$ ($\bigcirc $),
taken from Fig.\ref{fig1}, and HgBa$_{2}$CuO$_{4+\protect\delta }$
($\Box $) \protect\cite{hoferhg} with x derived from
Eq.(\ref{eq1}). The solid and dashed lines corresponds to Eq.(\ref
{eq2}) with $x_{0}=0.27$, $a_{T_{c}\protect\gamma }=5909\ \ K$ for La$_{2-x}$%
Sr$_{x}$CuO$_{4}$ and $a_{T_{c}\protect\gamma }=28830\ \ K$ for HgBa$_{2}$CuO%
$_{4+\protect\delta }$.} \label{fig2}
\end{figure}

\begin{figure}
\centering
\includegraphics[width=0.95\linewidth]{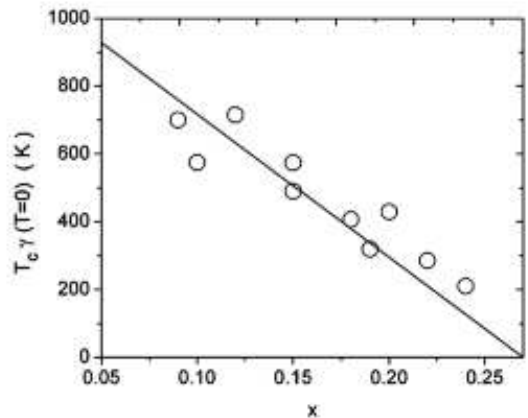} \vskip12pt
\caption{$\ T_{c}\protect\gamma _{T=0}$ versus dopant
concentration $x$ for La$_{2-x}$Sr$_{x}$CuO$_{4}$ ($\bigcirc $),
taken from from the penetration
depth data\protect\cite{shibauchi,panagopoulos} in terms of $\protect\gamma %
_{T=0}=\protect\lambda _{c}\left( T=0\right) /\protect\lambda
_{ab}\left( T=0\right) $. The solid line corresponds to
Eq.(\ref{eq3}) with $x_{0}=0.27$ and $a_{T_{c}\protect\gamma
}=4220\ \ K$ .} \label{fig3}
\end{figure}

Another essential property in this context is the anisotropy
parameter $\gamma $. In
tetragonal cuprates it is defined by $\gamma =\xi _{ab}/\xi _{c}$, because $%
\xi _{a}=\xi _{b}=\xi _{ab}$. In the normal state, where $\gamma
=\xi _{ab}/\xi _{c}=\sqrt{\rho _{ab}/\rho _{c}}$, it can be
inferred from resistivity\cite{takagi,sasagawa} and magnetic
torque\cite{hoferdis} measurements. In the superconducting state,
where $\gamma =\xi _{ab}/\xi _{c}=\lambda _{c}/\lambda _{ab}$
holds, it follows from magnetic torque\cite {hoferdis,hoferhg} and
penetration depth data\cite{shibauchi,panagopoulos} . $\lambda
_{ab}$ and $\lambda _{c}$ are the London penetration depth due to
supercurrents flowing parallel and perpendicular to the ab-planes,
respectively. In a mean-field treatment of the Ginzburg-Landau Hamiltonian $%
\gamma $ can also be expressed in terms of the effective masses
$M_{ab}$ and $M_{c\text{ }}$ for the motion of the pairs parallel
and perpendicular to the ab-planes as $\gamma
=\sqrt{M_{c}/M_{ab}}$. As shown in Fig.\ref{fig1} for
La$_{2-x}$Sr$_{x}$CuO$_{4}$, the anisotropy parameter $\gamma
_{T_{c}}$, evaluated close to $T_{c}$, increases as the dopant
concentration is reduced
and appears to diverge in the underdoped limit. This doping dependence of $%
\gamma _{T_{c}}$, also observed in HgBa$_{2}$CuO$_{4+\delta }$
\cite{hoferhg} and YBa$_{2}$Cu$_{3}$O$_{7-x}$ \cite{janossy}, is
thought to be generic and
uncovers the 3D-2D crossover, tuned by reducing the dopant concentration $x$%
. Consequently, the doping does not control $T_{c}$ only, but the
dimensional crossover, not included in 2D- models, as well. Since
$T_{c}$ tends to zero while $\gamma _{T_{c}}$ appears to diverge
in the underdoped limit, it is instructive to consider the doping
dependence of $T_{c}\gamma _{T_{c}}$ and $T_{c}\gamma _{T=0}$. A
glance to Figs.\ref{fig2} and \ref {fig3} makes it clear that
these quantities scale nearly linearly with the relative distance
from the overdoped limit, $x_{o}-x$. The straight lines correspond
to
\begin{equation}
T_{c}\gamma _{T_{c}} = a_{T_{c},\gamma }(x_{o}-x) \label{eq2}
\end{equation}
and
\begin{equation}
T_{c}\gamma _{T=0} = a_{T=0,\gamma }(x_{o}-x), \label{eq3}
\end{equation}
respectively. To check the consistency of the empirical relations (\ref{eq1}%
), (\ref{eq2}) and (\ref{eq3}) we included in Fig.\ref{fig1} the$\
T_{c}$ values, resulting from Eqs.(\ref{eq2}) and (\ref{eq3}), and
the experimental data for $\gamma $. Moreover, we included $\gamma
_{T_{c}}\left( x\right) $, expressed in terms of Eqs.(\ref{eq1})
and (\ref{eq2}).
Close to both, the
QSI and the QSN transition, the empirical relation (\ref{eq1}) for $%
T_{c} $ implies the power law behavior
\begin{equation}
T_{c} = 2T_{c}^{m}\sqrt{82.6}\left| x-x_{u,o}\right| , \label{eq4}
\end{equation}
and with the empirical doping dependence of $T_{c}\gamma _{T_{c}}$
(Eq.(\ref {eq2})) and $T_{c}\gamma _{T=0}$ (Eq.(\ref{eq3})) the
expressions:
\begin{equation}
\gamma _{T_{c}}=\frac{a_{T_{c},\gamma }{}(x_{0}-x)}{2T_{c}^{m}\sqrt{82.6}%
\left( x-x_{u}\right) },\ \ \gamma _{T=0}=\frac{a_{T=0,\gamma }{}(x_{0}-x)}{%
2T_{c}^{m}\sqrt{82.6}\left( x-x_{u}\right) }.  \label{eq5}
\end{equation}
Their divergence in the underdoped limit $\left( x=x_{u}\right) $
implies a 2D-QSI transition and for any $x_{u}<x\leq x_{0\text{
}}$, where $\gamma _{T_{c}}$ and $\gamma _{T=0}$ are finite, a
superconducting state with anisotropic but genuine 3D character.
Consequently, the available
experimental data for the doping dependence of $T_{c}$, $\gamma _{T=0}$ and $%
\gamma _{T_{c}}$ strongly suggest that two dimensional models
cannot explain the occurrence of superconductivity in the
cuprates. Despite this, the vast majority of theoretical models
focus on single Cu-O plane,i.e., on the limit of zero intracell
and intercell c-axis coupling.

Next we confirm and extend this empirical scenario by invoking the
scaling theory of quantum critical phenomena. This theory predicts
that close to the QSI and QSN\ transition $T_{c}\left( x\right) $
scales as \cite {book,klosters,kim}

\begin{equation}
T_{c}\left( x\right) =a_{QSIT_{c}}\ \left( x-x_{u}\right) ^{z_{QSI}\overline{%
\nu }_{QSI}}  \label{eq6}
\end{equation}
and
\begin{equation}
T_{c}\left( x\right) =a_{QSNT_{c}}\ \left( x_{0}-x\right) ^{z_{QSN}\overline{%
\nu }_{QSN}},  \label{eq7}
\end{equation}
respectively. $z$ is the dynamic critical exponent, $\overline{\nu
}$ the
exponent of the correlation lengths $\xi \left( T=0\right) \propto \delta ^{-%
\overline{\nu }}$, $a_{QSI}$ and $a_{QSN}$ denote nonuniversal
critical amplitudes. Moreover, supposing that the QSI transition
occurs in 2D, $T_{c}$ and the zero temperature in-plane
penetration depth ${\lambda _{ab}(T=0)}$ are not independent but
in the limit $x-x_{u}\rightarrow 0$ related by the universal
relation \cite{book,klosters,kim}
\begin{equation}
\frac{T{_{c}\lambda
_{ab}^{2}(T=0)}}{d_{s}}=\frac{1}{\overline{Q}_{2}}\left(
{\frac{\Phi _{0}^{2}}{16\pi ^{3}k_{B}}}\right) ,  \label{eq8}
\end{equation}
where $\overline{Q}_{2}$ is a universal number. Here the bulk
superconductor
corresponds to a stack of independent superconducting slabs of thickness $%
d_{s}$. Experimentally, this relation is well confirmed for
various families of underdoped cuprates in terms of the empirical
Uemura plot \cite {book,klosters,uemura}. On the other hand, at
finite temperature there is along the 3D phase transition line
$T_{c}\left( x\right) $ the universal relation\cite{book,tsprb},
\begin{equation}
k_{B}T_{c}=\frac{\Phi _{0}^{2}}{16\pi ^{3}}\frac{\xi
_{ab,0}^{-}}{\gamma _{T_{c}}\lambda _{ab,0}^{2}}.  \label{eq9}
\end{equation}
$\xi _{ab,0}^{-}$ and $\lambda _{ab,0}$ are the finite temperature
critical amplitudes of in-plane correlation length for $T\leq
T_{c}$ and penetration depth, respectively. Matching with the
quantum behavior (\ref{eq6}) and (\ref
{eq8}) requires that the finite temperature critical amplitudes scale as, $%
\xi _{ab,0}^{-}\propto \xi _{ab}\left( T=0\right) \propto \left(
x-x_{u}\right) ^{-\overline{\nu }_{QSI}}$ and $\lambda _{ab,0}^{-2}\propto {%
\lambda _{ab}^{-2}(0)\propto }\left( x-x_{u}\right) ^{z_{QSI}\overline{\nu }%
_{QSI}}$. Substitution into Eq.(\ref{eq9}) leads to the remarkable
result that close to the 2D-QSI transition $\gamma _{T_{c}}$
diverges as
\begin{equation}
\gamma _{T_{c}} \propto \left( x-x_{u}\right) ^{-\overline{%
\nu }_{QSI}}.  \label{eq10}
\end{equation}
On the other hand, approaching the 2D-QSI transition at $T=0$
along the doping axis, the singular part of the ground state
energy density scales in 3D as \cite{book,kim}
$E_{s}^{3D}=\overline{R}_{3}\left( \xi _{\tau }\xi _{ab}^{2}\xi
_{c}\right) ^{-1}=\overline{R}_{3}\gamma _{T=0}\left( \xi
_{\tau }\xi _{ab}^{3}\right) ^{-1}$, while in 2D, $E_{s}^{2D}=\overline{R}%
_{2}\left( \xi _{\tau }\xi _{ab}^{2}d_{s}\right) ^{-1}$ applies. $\overline{R%
}_{3}$ and $\overline{R}_{2}$ are universal numbers and $\xi
_{\tau }\propto \left( x-x_{u}\right) ^{-z\overline{\nu }}$ is the
temporal correlation length having units of energy. The 3D-2D
crossover requires that both expressions give close to the QSI
transition the same doping dependence. This yields
\begin{equation}
\gamma _{T=0} =\overline{R}_{3}\xi _{ab}/\left( \overline{R}%
_{2}d_{s}\right) \propto \left( x-x_{u}\right) ^{-\overline{\nu
}_{QSI}}, \label{eq11}
\end{equation}
in agreement with Eq.(\ref{eq10}). Accordingly, given a 2D-QSI
transition we confirmed the divergence of $\gamma _{T_{c}}$ and
$\gamma _{T=0}$ , emerging from the empirical scenario (
Eq.(\ref{eq5})). Moreover, comparing Eqs.(\ref {eq4})-(\ref{eq6}),
(\ref{eq10}) and (\ref{eq11}) this scenario also
suggests a 2D-QSI transition with critical exponents $z_{QSI}=1$ and $%
\overline{\nu }_{QSI}=1$. Although the experimental data are
rather sparse close to this transition, there is mounting evidence
for $z_{QSI}\approx 1$ and $\overline{\nu }_{QSI}\approx
1$\cite{book,klosters,housten,tshk,tsprb}. These estimates are
close to theoretical predictions \cite{fisher,herbut}, from which
$z_{QSI}=1$ is expected for a disordered bosonic system with
long-range Coulomb interactions independent of dimensionality and $\overline{%
\nu }_{QSI}\geq 1\approx 1.03$ in D=2. They are also consistent
with Monte
Carlo calculations on the dirty-boson Hamiltonian, yielding $\overline{\nu }%
_{QSI}=0.9\pm 0.15$\cite{wallin}, as well as with experiments on
the 2-QSI transition in thin (InO, Bi, MoGe) films, where
$\overline{\nu }_{QSI}$ clusters around $\overline{\nu
}_{QSI}=1.2\pm 0.2$\cite {hebard,yazdani,markovic}. In this
transition, the loss of phase coherence is due to the localization
of the pairs which is ultimately responsible for
the transition. As the 3D-QSN transition is concerned, the estimate for $%
z_{QSN}\overline{\nu }_{QSN}$ , inferred from the empirical
relation (\ref {eq4}) is $1$. This value is consistent with the
disordered d-wave superconductor to normal state transition
considered by Herbut \cite
{herbutd}, with $z_{QSN}\overline{\nu }_{QSN}\approx 1,\ z_{QSN}=2$ and $%
\overline{\nu }_{QSN}\approx 1/2$. Upon using Eqs.(\ref{eq6}),
(\ref{eq10}) and (\ref{eq11}) one obtains for $T_{c}\left(
x\right) \gamma _{T_{c}}\left( x\right) $ and $T_{c}\left(
x\right) \gamma _{T=0}\left( x\right) $ the expression
\begin{equation}
T_{c} \gamma _{T_{c}} \propto
T_{c} \gamma _{T=0} \propto \left( x-x_{u}\right) ^{-%
\overline{\nu }_{QSI}\left( z_{QSI}\ -1\right) }.  \label{eq12}
\end{equation}
For $z_{QSI}=1$ these quantities remain finite, as suggested by
the experimental data shown in Figs.\ref{fig2} and \ref{fig3}.
More importantly, eliminating $x-x_{u}$ from the scaling forms
(\ref{eq6}), (\ref{eq8}), (\ref {eq10}) and (\ref{eq11}) we obtain
close to the 2D-QSI transition the novel scaling form
\begin{equation}
T_{c} \propto n_{s}\left( T=0\right) \propto \gamma _{T_{c}}\left(
x\right) ^{-z_{QSI}}\propto \gamma _{T=0}^{-z_{QSI}}, \label{eq13}
\end{equation}
where $n_{s} \left( T=0,x\right) \propto d_{s}/\lambda
_{ab}^{2}\left( T=0,x\right) $ is the aerial superfluid number
density. It relates the superconducting properties to the
anisotropy parameters, fixing the dimensionality of the system,
for any $z_{QSI}$. Since $z_{QSI}>0$, this leads to the conclusion
that superconductivity in the cuprates oxides is unalterably
linked to a finite $\gamma _{T=0}$,\ implying an anisotropic but
3D condensation mechanism.

Finally we note that in the regime where thermal fluctuations dominate, $%
T_{c}\gamma _{T_{c}}$ is proportional to the number density
$\left\langle \left| \Psi \right| ^{2}\right\rangle $ of the pairs
at $T_{c}$. The complex order parameter field $\Psi $ represents
the wavefunction of the pairs of charge 2e. In the mean-field
approximation, where $\left\langle \left| \Psi \right|
^{2}\right\rangle =\left\langle \left| \Psi \right| \right\rangle
^{2}$, this quantity vanishes for $T\geq T_{c}$. Thus, the plot $T_{c}$ $%
\gamma _{T_{c}}$ versus $x$, shown in Fig.\ref{fig2}, also reveals
the existence of pairs at and above $T_{c}$, as well as a raise of
their number density with decreasing dopant concentration.

In conclusion, we have shown that the scenario emerging from the
empirical doping dependence of $T_{c}$, $\gamma _{T_{c}}$ and
$\gamma _{T=0}$ is remarkably consistent with the scaling
properties resulting from a 3D critical line $T_{c}\left( x\right)
$, ending in the underdoped and overdoped limit. Here the doping
tuned QSI and 3D-QSN transitions occur. Given the evidence for a
2D-QSI transition and the associated 3D-2D crossover, tuned by
decreasing dopant concentration, we derived the scaling relation
(\ref{eq13}), relating the essential characteristics of the
superconducting phase to the anisotropy parameters, fixing the
dimensionality of the system. It implies that for any finite value
of the dynamic critical exponent ($z_{QSI}$) of the quantum
superconductor to insulator transition, a non vanishing transition
temperature and superfluid aerial superfluid density ($n_{s}
\left( T=0,x\right) $) in the ground state, require a finite
anisotropy. The important conclusion there is that a non vanishing
superfluid density in the ground state of the cuprates oxides is
unalterably linked to an anisotropic but 3D condensation
mechanism. This finding clearly reveals that two dimensional
models cannot explain the occurrence of superconductivity in the
cuprates. They single out, however, theories that ascribe the
phenomenon to interlayer coupling
\cite{bulaevskii,tsgceur,tsgczphys,andersonint}.
\bigskip

The author is grateful to H. Keller, K.A. M\"{u}ller and S. Roos
for very useful comments and suggestions on the subject matter.

\end{multicols}

\end{document}